\documentclass[aps,prl,showpacs,groupedaddress,superscriptaddress,twocolumn,secnumarabic,nobibnotes]{revtex4}
\usepackage{amsmath,graphicx,bm,xcolor}

\begin{document}
%%%%%%%%%%%%%%%%%%%%%%%%%%%%%%%%%%%%%%%%%%%%%%%%%%%%%
\title{Configuration-Sensitive Transport on Domain Walls of a Magnetic Topological Insulator}
%%%%%%%%%%%%%%%%%%%%%%%%%%%%%%%%%%%%%%%%%%%%%%%%%%%%%
\author{Yan-Feng Zhou}
\affiliation{International Center for Quantum Materials, School of Physics, Peking University, Beijing 100871, China}
\affiliation{Collaborative Innovation Center of Quantum Matter, Beijing 100871, China}

\author{Zhe Hou}
\affiliation{International Center for Quantum Materials, School of Physics, Peking University, Beijing 100871, China}
\affiliation{Collaborative Innovation Center of Quantum Matter, Beijing 100871, China}

\author{Qing-Feng Sun}
\email[]{sunqf@pku.edu.cn}
\affiliation{International Center for Quantum Materials, School of Physics, Peking University, Beijing 100871, China}
\affiliation{Collaborative Innovation Center of Quantum Matter, Beijing 100871, China}
\affiliation{CAS Center for Excellence in Topological Quantum Computation, University of Chinese Academy of Sciences, Beijing 100190, China}

\date{\today}
%%%%%%%%%%%%%%%%%%%%%%%%%%%%%%%%%%%%%%%%%%%%%%%%%%%%%
\begin{abstract}
We study the transport on the domain wall (DW) in a magnetic topological insulator.
The low-energy behaviors of the magnetic topological insulator are dominated
by the chiral edge states (CESs).
Here, we find that the spectrum and transport of the CESs at the DW are strongly
dependent on the DW configuration.
For a Bloch wall, two co-propagating CESs at the DW are doubly degenerate
and the incoming electron is totally reflected.
However, for a N\'{e}el wall, the two CESs are split
and the transmission is determined by the interference between the CESs.
Moreover, the effective Hamiltonian for the CESs indicates that
the component of magnetization perpendicular to the wall leads to the distinct transport behavior. These findings may pave a way to realize the low-power-dissipation spintronics
devices based on magnetic DWs.
\end{abstract}

\maketitle
%%%%%%%%%%%%%%%%%%%%%%%%%%%%%%%%%%%%%%%%%%%%%%%%%%%%%%

\paragraph*{Introduction.}
The discovery of topological insulator (TI) has attracted intensive interest
in searching for topologically non-trivial states of condensed matter and subsequently,
triggered a series of occurrences of novel physical effects\cite{HMZ,Qxl}.
The quantum anomalous Hall effect (QAHE), i.e.,
quantum Hall effect without the external magnetic field,
can be achieved in magnetic TI by introducing ferromagnetism in TI.\cite{LiuCX,YuR}
The magnetic TI has an insulating bulk classified by a Chern number $\mathcal{C}$
and $\mathcal{C}$ conducting chiral edge states (CESs) through bulk-boundary correspondence.
In recent, QAHE has been
experimentally realized in Cr-doped\cite{ChangCZ1,Checkelsky,Kou,Bestwick,Kandala} and V-doped\cite{ChangCZ2} $\mathrm{(Bi,Sb)_2Te_3}$ magnetic TI thin films,
and the Hall resistance shows a quantized value
$\pm h/e^2$ implying that the Chern number of the
magnetic TIs $\mathcal{C}=\pm1$ which can be controlled by the magnetization direction\cite{LiuCX2}.

The boundary between magnetic TI domains of opposite magnetization
with $\mathcal{C}=\pm1$ forms a magnetic domain wall (DW) with
a magnetization rotation to minimize the total magnetic energy as shown in Fig.1(a).
Both the optimized configuration and thickness of the DW are determined by a balance
between competing energy contributions\cite{book1,book2}.
Two energetically favorable configurations are Bloch wall and N\'{e}el wall,
and the DW configuration can be controlled by Dzyaloshinskii-Moriya interaction\cite{Thiaville,ChenG1,ChenG2,DeJong}.
Moreover, due to the different chirality of CESs across the DW, two
co-propagating CESs are expected to reside on the DW.
Very recently, the DWs of magnetic TI have been realized in Cr-doped
$\mathrm{(Bi,Sb)_2Te_3}$ by the tip of a magnetic force microscope\cite{YasudaK}
and by spatially modulating the external magnetic
field using Meissner repulsion from a bulk superconductor\cite{Rosen},
and the chiral transport of CESs has been observed in these experiments.
Owing to the robustness of the CESs against backscattering, the DWs of magnetic TI have potential applications in the low-power-consumption spintronic devices, such as the nonvolatile racetrack memory\cite{Parkin}.

\begin{figure}
\includegraphics[scale=0.31]{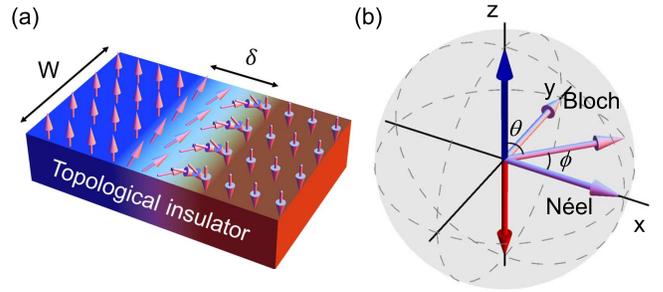}
\caption{
(a) Schematic diagram of a magnetic DW between two magnetic TI domains.
The direction of magnetization vector $\bm{\mathrm{M}}$ rotates continuously
from $+z$ to $-z$ direction inside the DW.
(b) The sphere of possible $\bm{\mathrm{M}}$ with the magnetization configurations corresponding to the different rotation modes defined by the azimuthal angle $\phi$.
Here $\phi=0$ and $\pi/2$ corresponds to N\'{e}el wall and Bloch wall, respectively.}
\end{figure}

In this Letter, we study the transport of a two-terminal device
containing a DW of thickness $\delta$ and width $W$ in a magnetic TI [see Fig.1(a)].
In the low energy case, the transport behaviors of the magnetic TI are dominated
by CESs at the device edges as well as at the DW.
We calculate the band structure of magnetic TI with both Bloch wall and N\'{e}el wall.
For Bloch wall, two co-propagating linear CESs at the DW are doubly degenerate,
while for N\'{e}el wall a split is present.
As a result, the transport property is strongly dependent on the DW configuration.
In the Bloch wall case, the incoming electron with zero energy is totally reflected
regardless of the system parameters.
However, in the N\'{e}el case, the device functions as a chirality-based
Mach-Zehnder interferometry, so that
the transmission coefficient oscillates between zero and unity with changes in system parameters.
By constructing the scattering matrix of the device from the effective Hamiltonian,
these transport behaviors can be well understood.

\paragraph*{Model.}
As shown in Fig.1(a), two magnetic TI domains with upwards (blue region)
and downwards (red region) magnetization are separated by a DW.
The magnetization vectors are homogeneous away from the DW and change
continuously from $+z$ direction to $-z$ direction inside the DW.
The configuration of the DW can be described by magnetization vector
$\bm{\mathrm{M}}(x)=(M_x,M_y,M_z)=M(\sin\theta\cos\phi, \sin\theta\sin\phi, \cos\theta)$,
with a constant magnitude $M$.
The azimuthal angle $\theta$ is a function of $x$ with $\cos\theta(x)=-\tanh\frac{x}{\delta}$
and the azimuthal angle $\phi$ defines the type of the magnetic DW.
From the sphere of possible magnetization vectors, the magnetic vector $\phi=0$
in N\'{e}el wall and $\phi=\pi/2$ in Bloch wall [see Fig.1(b)].

The low-energy states of magnetic TI can be described by the Hamiltonian\cite{YuR,WangJ1} $H=\sum_{\bm{k}}\Psi_{\bm{k}}^{\dag}H(\bm{k})\Psi_{\bm{k}}$ with
\begin{equation}\label{eq1}
  H(\bm{k})=
  \nu_Fk_y\sigma_x\tau_z-\nu_Fk_x\sigma_y\tau_z+m(\bm{k})\tau_x+\bm{M}\cdot\bm{\sigma},
\end{equation}
where the momentum $\bm{k}=(k_x, k_y)$ and $\Psi_{\bm{k}}=[\psi_{t\uparrow},\psi_{t\downarrow},\psi_{b\uparrow},\psi_{b\downarrow}]^T$
being a four component electron operator,
where $t$ and $b$ label electrons from the top and bottom layers,
and $\uparrow$ and $\downarrow$ denote electrons with spin up and down, respectively.
$\sigma_{x,y,z}$ and $\tau_{x,y,z}$ are Pauli matrices for spin and layer.
$m(\bm{k})=m_0-m_1(k_x^2+k_y^2)$ describes the coupling between the top and bottom layer.
In the calculation, we set the Fermi velocity $\nu_F=0.222\ \mathrm{eVnm}$,
$m_0=0.026\ \mathrm{eV}$, $m_1=0.137\ \mathrm{eVnm^2}$, and $M=0.048\ \mathrm{eV}$.\cite{WangJ2}
As $M>m_0$, the magnetic TIs with Chern number $\mathcal{C}=\pm1$ are realized in the domains
with homogeneous upwards and downwards magnetization.
For the numerical calculation, we discretize the Hamiltonian Eq.(\ref{eq1}) into a lattice version with a lattice constant $a=0.6\ \mathrm{nm}$.\cite{SDatta,ChenCZ1,ChenCZ2,LiYH}

\begin{figure}
\includegraphics[scale=0.3]{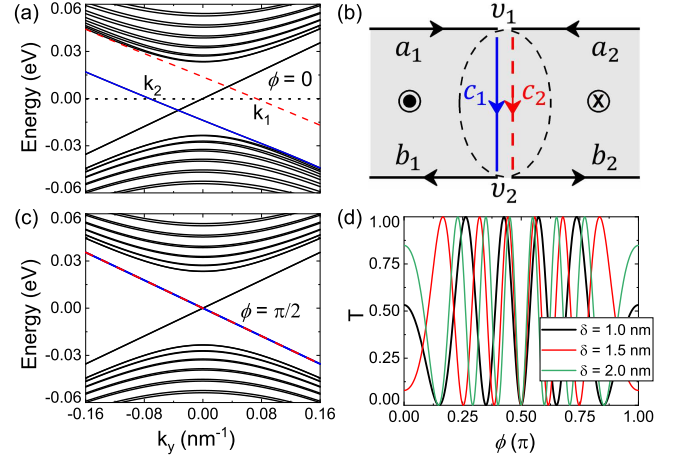}
\caption{
(a) and (c) The band structure of an infinite slab of magnetic TI extending along the $y$
direction with a N\'{e}el wall ($\phi=0$) in (a) and Bloch wall ($\phi=\pi/2$) in (c).
The width of the slab is $180.6\ \mathrm{nm}$ and the thickness of the DW is $1\ \mathrm{nm}$.
The blue solid and red dashed lines represent the chiral modes on the DW.
(b) Schematic depicting the transport process based on the chiral modes.
(d) The zero-energy transmission coefficient $T$ of the device in Fig.1(a) versus $\phi$
for several DW thickness $\delta$ with the width $W=90\ \mathrm{nm}$.
}
\end{figure}

\paragraph*{Chiral modes on the DW.}
To study the spectrum of the CESs, we first consider an infinite slab of magnetic TI
containing a DW [see Fig.1(a)] which extends along the $y$ direction
and has a finite width in $x$ direction.
As the slab is invariant by translating along the $y$ axis,
the momentum $k_y$ is a good quantum number.
Figures 2(a) and 2(c) show the band structure of the slab
with a N\'{e}el wall ($\phi=0$) and Bloch wall ($\phi=\pi/2$), respectively.
Inside the bulk gap, there are four linear chiral modes with
two co-propagating modes along the DW (blue solid and red dashed lines)
and two degenerate modes along the slab edges propagating in opposite direction
(black solid lines).
The presence of two chiral modes residing on the DW arises from the change
in Chern number from +1 to -1 across the DW.
For Bloch wall, the co-propagating modes on the DW are degenerate,
while for N\'{e}el wall, the chiral modes are split with energy
dispersions $E \propto -k_y\pm \Delta k/2$.
As the DW is located inside the slab, it has no effects on the chiral modes
on the edges as shown in Fig.2(a) and (c).

Let us construct the one-dimensional effective Hamiltonian for the co-propagating chiral modes on the DW to make the split clear. By a unitary transformation
\begin{equation}\label{eq2}
U=\frac{1}{\sqrt{2}}\left(
                \begin{array}{cccc}
                  1 & 0 & 1 & 0 \\
                  0 & 1 & 0 & -1 \\
                  0 & 1 & 0 & 1 \\
                  1 & 0 & -1 & 0 \\
                \end{array}
              \right),
\end{equation}
the Hamiltonian (\ref{eq1}) becomes
\begin{equation}\label{eq3}
H'(k)=\left(\begin{matrix}
  H_+ & M_{\|} \\
           M_{\|}^{\dag} & H_-
\end{matrix}\right),
\end{equation}
with
\begin{equation}\label{eq4}
H_{\pm}=\nu_Fk_y\tilde{\sigma}_x\mp\nu_Fk_x\tilde{\sigma}_y+(m(k)\pm M_z)\tilde{\sigma}_z,
\end{equation}
in terms of new basis $(\psi_{+\uparrow},\psi_{-\downarrow},\psi_{+\downarrow},\psi_{-\uparrow})^T$ with
$\psi_{\pm\uparrow}=(\psi_{t\uparrow}\pm\psi_{b\uparrow})/\sqrt{2}$ and
$\psi_{\pm\downarrow}=(\psi_{t\downarrow}\pm\psi_{b\downarrow})/\sqrt{2}$, and $M_{\|}=M_x-iM_y\sigma_z$. $\tilde{\sigma}_{x,y,z}$ are Pauli matrices.
Inside the DW with magnetization vector $\bm{\mathrm{M}}(x)=M(\mathrm{sech}\frac{x}{\delta}\cos\phi, \mathrm{sech}\frac{x}{\delta}\sin\phi,-\tanh\frac{x}{\delta})$,
both $H_+$ and $H_-$ are nontrivial due to the sign change of $M_z$ across the DW,
so that there exist two chiral states\cite{QiXL2,ZhangRX}.
As $H_\pm$ are coupled by element $M_{\|}$ in Eq.(\ref{eq3}),
to find the solutions of chiral states, we replace $k_x\rightarrow-i\partial_x$
and decompose the Hamiltonian as $H'=H_0+\Delta H$,
in which $H_0$ contains the decoupled $H_\pm$ and $\Delta H$ consists of the element $M_{\|}$.
We solve $H_0$ first and treat $\Delta H$ as a perturbation\cite{SShen,WangJ3}.

First, we solve the eigenequation $H_+\zeta_+(x)=E\zeta_+(x)$ for $k_y=0$ and $E=0$.
It can be checked that $H_+(k_y=0)$ and $\tilde{\sigma}_x$ satisfy the anticommutation relation
$\{H_+(k_y=0),\tilde{\sigma}_x\}=0$.
Thus, the zero-energy eigenstate is the simultaneous eigenstate of $H_+$ and $\tilde{\sigma}_x$.
Consider the ansatz $\zeta_+ (x)=\eta_+^s(x)\chi_x^s$, where $\tilde{\sigma}_x\chi_x^s=s\chi_x^s$, $s=\pm$, we have
\begin{equation}\label{Hp1}
  (s\nu_F\partial_x+m_0+m_1\partial_x^2+M_z)\eta_+^s(x) = 0 .
\end{equation}
With a substitution $u=(1+e^{2x/\delta})^{-1}$,\cite{Flugge} we arrive at the hypergeometric form of Eq.(\ref{Hp1}) and find the solution $\zeta_+(x)= \eta_+^-(x)\chi_x^-=
K_1u^\alpha(1-u)^\beta \ _{2}F_1(\alpha+\beta,\alpha+\beta+1,2\alpha+1+\frac{\delta\nu_F}{2m_1};u)(1,-1)^T$ with
(see \cite{Supp} for details)
\begin{eqnarray*}
  \alpha &=& \frac{\delta}{2}\frac{\sqrt{\nu_F^2-4m_1(m_0-M)}-\nu_F}{2m_1}, \\
  \beta &=& \frac{\delta}{2}\frac{\nu_F-\sqrt{\nu_F^2-4m_1(m_0+M)}}{2m_1}.
\end{eqnarray*}
Similarly, we find the solution of $H_-$,
$\zeta_-(x)=\eta_-^-(x)\chi_x^- =
K_2g^\alpha(1-g)^\beta \ _{2}F_1(\alpha+\beta,\alpha+\beta+1,2\alpha+1+\frac{\delta\nu_F}{2m_1};g)(1,-1)^T$ with $g=(1+e^{-2x/\delta})^{-1}$.
Here $K_{1,2}$ is normalization factor and $_2F_1$ is the hypergeometric function.
Written in a four-component notation, $\zeta_+(x)=\eta_+^-(x)(1,-1,0,0)^T$ and $\zeta_-(x)=\eta_-^-(x) (0,0,1,-1)^T$.

Next, we consider the perturbation term $\Delta H$ by projecting the Hamiltonian $H'(k)$ onto the two zero-energy states leading to the one-dimensional effective Hamiltonian\cite{SShen,WangJ3},
\begin{equation}\label{effect}
  H_\mathrm{eff}=\left(
                     \begin{array}{cc}
                       \langle\zeta_+|H'|\zeta_+\rangle & \langle\zeta_+|H'|\zeta_-\rangle \\
                       \langle\zeta_-|H'|\zeta_+\rangle & \langle\zeta_-|H'|\zeta_-\rangle \\
                     \end{array}
                   \right).
\end{equation}
It can easily be obtained that $\langle\zeta_+|H'|\zeta_+\rangle=\langle\zeta_-|H'|\zeta_-\rangle=-\nu_Fk_y$
and the nondiagonal element depends on the type of the DW.
For N\'{e}el wall, the magnetization vector $\bm{\mathrm{M}}(x)=M(\mathrm{sech}\frac{x}{\delta}, 0,-\tanh\frac{x}{\delta})$, so $M_{\|}=M\mathrm{sech}\frac{x}{\delta}\mathbf{I}_{2\times2}$
with the $2\times2$ unit matrix $\mathbf{I}_{2\times2}$. The effective Hamiltonian becomes
\begin{equation}\label{neel}
H_{\mathrm{N\acute{e}el}}(k_y)=\left(\begin{matrix}
  -\nu_Fk_y & \kappa\\
           \kappa^{\ast} & -\nu_Fk_y
\end{matrix}\right),
\end{equation}
where $\kappa =\int \eta_+^{-\ast}(x)M\mathrm{sech}\frac{x}{\delta} \eta_-^-(x)dx$
is the hybridization of the two states.
The excitation spectrum is $E(k_y)=-\nu_Fk_y\pm |\kappa|$.
These two modes are the nondegenerate chiral modes with a splitting
$\Delta k =k_1-k_2= 2|\kappa|/\nu_F$ in $k_y$ [blue solid and red dashed lines in the Fig.2(a)].
However, for Bloch wall, $\bm{\mathrm{M}}(x)=M(0,\mathrm{sech}\frac{x}{\delta}, -\tanh\frac{x}{\delta})$, so $M_{\|}=-iM\mathrm{sech}\frac{x}{\delta}\sigma_z$ and $\langle\zeta_+|H'|\zeta_-\rangle=0$. The effective Hamiltonian becomes
\begin{equation}\label{bloch}
H_{\mathrm{Bloch}}(k_y)=\left(\begin{matrix}
  -\nu_Fk_y & 0\\
           0 & -\nu_Fk_y
\end{matrix}\right).
\end{equation}
The excitation spectrum is doubly degenerate with $E(k_y)=-\nu_Fk_y$ in accordance with Fig. 2(c). At this point, it can be seen that the split between the co-propagating chiral modes results from the $x$ component of the magnetization inside the DW and $\Delta k$ depends
on the type and thickness of the DW.

\paragraph*{Transport on the DW.}
To study the effect of DW configuration on the transport of the DW of magnetic TI,
we construct a two-terminal device [see Fig.1(a)] which contains a DW in the center region
and two semi-infinite left and right magnetic TI domains.
For low incident energy, the transport occurs via the CESs and
Fig.2(b) depicts the transport process.
By using the
nonequilibrium Green's function method, the transmission coefficients can be
obtained from\cite{SDatta,sun1,sun2,sun4} $T(E)=\mathrm{Tr}[\Gamma_LG^r\Gamma_RG^a]$,
with the incident energy $E$, retarded/advanced Green's function $G^{r/a}(E)$, and
line-width function $\Gamma^{L/R}(E)$.

\begin{figure}
\includegraphics[scale=0.3]{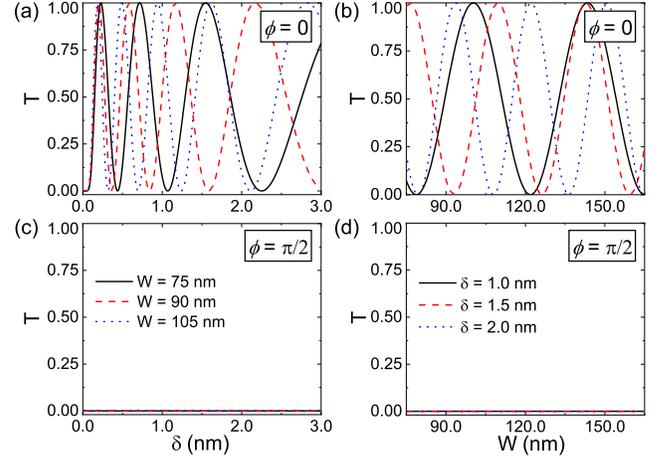}
\caption{(a) and (c) The transmission coefficient $T$ versus DW thickness $\delta$ for
(a) N\'{e}el wall ($\phi=0$) and for (c) Bloch wall ($\phi=\pi/2$) in several widths $W$ for $E=0$. (b) and (d) $T$ versus width $W$ for (b) N\'{e}el wall and (d) Bloch wall in several DW thicknesses $\delta$ for $E=0$.}
\end{figure}

When an electron propagating along the mode $a_1$ (black arrow from
the left terminal) arrives at the trijunction $\nu_1$,
it is scattered into the chiral modes $c_1$ and $c_2$ in the DW region as shown in Fig.2(b).
After the propagation along the DW, the electron is scattered
off the trijunction $\nu_2$ and gets into the outgoing modes $b_1$ and $b_2$ eventually.
Fig.2(d) shows the transmission coefficient $T$ at $E=0$
as a function of $\phi$ which specifies the type of the DW.
$T$ is the periodic function of $\phi$ with the period $\pi$,
so we only show the results for $0\leq\phi\leq\pi$.
It can be observed in Fig.2(d) that for Bloch wall ($\phi=\pi/2$),
the transmission coefficient $T=0$ and remains unchanged with the change
in the DW thickness $\delta$.
However, deviating from $\phi=\pi/2$, $T$ oscillates between 0 and 1 with the change
in $\phi$ and DW thickness $\delta$, and is symmetric about $\phi=\pi/2$,
i.e. $T(\phi) = T(\pi-\phi)$.
These results suggest that the current of the device in Fig.1(a)
can be switched on or off by changing the magnetization configuration of the DW.
Such a switch effect has an underlying application in spintronics, because that
the current is completely layer-locked spin-polarized\cite{ZhangRX,WuJ}.

Let us study the N\'{e}el wall and Bloch wall in detail.
Fig.3 shows the dependence of transmission coefficient $T$
on the DW thickness $\delta$ and device width $W$.
For N\'{e}el wall, $T$ approaches zero as the thickness of the DW vanishes [see Fig.3(a)].
With increasing in the thickness of the DW, $T$ oscillates between 0 and 1 for a fixed width $W$.
The thinner the DW is, the faster $T$ oscillates.
Moreover, $T$ shows a periodic function of the device width $W$
and the period is small for thick DW [see Fig.3(b)].
These imply that the device with N\'{e}el wall exhibits the behavior of a two-path interferometer.
However, for Bloch wall, the transmission coefficient $T$ is vanishing regardless of
the system parameters [see Fig.3(c,d)].
At this point, we can see that the two different DWs show absolutely different transport behaviors.
Below, based on the effective Hamiltonian Eqs.(\ref{neel} and \ref{bloch}), we construct the scattering matrix $S$ of the two-terminal device to understand the underlying physics.

To find the scattering matrix which relates the incoming modes to the outgoing modes,
we return to the Hamiltonian $H'(k)$ [see Eqs.(\ref{eq3} and \ref{eq4})]
to see the origin of the chiral modes $a_{1,2}$ and $b_{1,2}$ in Fig.2(b).
For left magnetic TI domain with $\bm{\mathrm{M}}=(0,0,M)$,
$M_{\|}=0$, $H_+$ is nontrivial and $H_-$ is trivial.
So $a_1$ and $b_1$ at the edge can be obtained by solving the Hamiltonian $H_+$
with open boundary conditions solely, which is similar with the mode $\zeta_+$.
On the other hand, for right magnetic TI domain with $\bm{\mathrm{M}}=(0,0,-M)$,
$H_+$ is trivial and $H_-$ is nontrivial.
Similarly, $a_2$ and $b_2$ can be obtained from the Hamiltonian $H_-$, which is similar with  $\zeta_-$.
Considering that $a_1,b_1$ and $\zeta_+$ ($a_2,b_2$ and $\zeta_-$) are the bound state solutions of the $H_+$ ($H_-$) and have the same chirality, at the trijunction $\nu_1$ [see Fig.2(b)],
the mode $a_1$ ($a_2$) is scattered onto $\zeta_+$ ($\zeta_-$) and at the trijunction $\nu_2$,
the mode $\zeta_+$ ($\zeta_-$) is scattered into $b_1$ ($b_2$).

For N\'{e}el wall, the solutions of the chiral modes on the DW [see Fig.2(b)] can be found as $c_{1,2}=\frac{1}{\sqrt{2}}(\zeta_+\pm\zeta_-)$
from the Hamiltonian $H_{\mathrm{N\acute{e}el}}$ in Eq.(\ref{neel}).
Thus, the scattering matrix of the trijunction $\nu_1$, $S_{\nu_1}=\frac{1}{\sqrt{2}}(\sigma_x+\sigma_z)$ accounts for the scattering of the incoming modes $a_{1,2}$ onto $c_{1,2}$.
Similarly, the scattering matrix describes the trijunction $\nu_2$ is $S_{\nu_2}=S_{\nu_1}$,
where the modes $c_{1,2}$ are scattered onto the outgoing modes $b_{1,2}$.
The scattering amplitude of the two-terminal device is found by
composing the scattering matrices,
\begin{equation}\label{6}
  S=S_{v_2}\left(
             \begin{array}{cc}
               e^{ik_1W} & 0 \\
               0 & e^{ik_2W} \\
             \end{array}
           \right)S_{v_1},
\end{equation}
where the second matrix contains the contribution of the dynamical phase and $k_{1,2}$ is the momentum of modes $c_{1,2}$.
In this case, the incoming electron from the chiral mode $a_1$ is equally split into
CESs $c_{1}$ and $c_{2}$ at $\nu_1$, then $c_{1,2}$ converge at $\nu_2$ and are finally scattered onto the
outgoing modes $b_{1,2}$, which serves as a Mach-Zehnder interferometry.\cite{Mach1,Mach2}
From Eq.(\ref{6}), the transmission coefficient is obtained as $T=\sin^2(\Delta kW/2)$
with $\Delta k=k_1-k_2$.
Fig.4 shows $\Delta k$ and $\sin^2(\Delta kW/2)$ as functions of the thickness $\delta$ of the DW. It can be seen that $\sin^2(\Delta kW/2)$ shows a good consistency with the $T$ of Fig.3(a)
and is a periodic function of the width of the device
in accordance with Fig.3(b).
Moreover, for a general DW defined by $\phi$, the hybridization $\kappa\propto\cos\phi$ so that the coefficient $\sin^2(\Delta kW/2)$ is the same for $\phi$, $\pi-\phi$, and $\pi +\phi$ [see Fig.2(d)].

\begin{figure}
\includegraphics[scale=0.3]{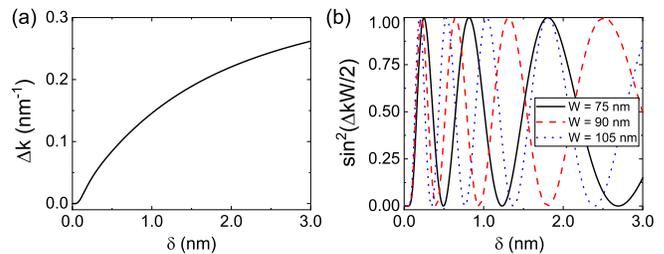}
\caption{(a) Momentum difference $\Delta k$ between the modes $c_{1,2}$ as a function of the thickness $\delta$ of the DW for zero energy extracted from the band structure [see Fig.2(a)].
(b) $\sin^2(\Delta kW/2)$ versus the thickness $\delta$ for several width $W$.}
\end{figure}

For Bloch wall, the co-propagating chiral modes on the wall are doubly degenerate
and $c_{1,2}=\zeta_\pm$ which can be obtained from the Hamiltonian $H_{\mathrm{Bloch}}$
in Eq.(\ref{bloch}).
This means that the incoming mode $a_1$ ($a_2$) is totally reflected onto $b_1$ ($b_2$).
This results a zero transmission coefficient which is consistent with Fig.3(c) and (d).
At this point, we have well understood the low-energy transport behavior of
the device containing a DW based on the effective Hamiltonian.

\paragraph*{Conclusions.}
In short, we find that the spectrum of the chiral modes is strongly dependent
on the detailed configuration of the DW.
For Bloch walls, the chiral modes are doubly degenerate,
while for N\'{e}el walls a split is present.
Correspondingly, the devices with different DW configuration show
very distinct transport behaviors.
In Bloch case, the current through the device vanishes regardless of system parameters.
However, in the N\'{e}el case, the transmission coefficient of the DW oscillates
between zero and unity with changes in system parameters and
is determined by the interference between the chiral modes.
From the scattering matrix of the device derived from the effective Hamiltonian of the chiral modes, these transport behaviors can be well understood.
These findings may pave a way to control the layer-locked spin-polarized current based on magnetic DWs.

\paragraph*{Acknowledgments.}
This work was financially supported by National Key R and D Program of China (2017YFA0303301),
NBRP of China (2015CB921102), and NSF-China (Grant No. 11574007).

\clearpage
\begin{widetext}
\begin{center}
  \bf{Supplemental Material for ``Configuration-Sensitive Transport on Domain Walls of a Magnetic Topological Insulator''}
\end{center}

\begin{center}
Yan-Feng Zhou, Zhe Hou, and Qing-Feng Sun
\end{center}

\section{Analytical solution of the Hamiltonian $H_\pm$}

First, we solve the eigenequation $H_+\zeta_+(x)=E\zeta_+(x)$ for $k_y=0$ and $E=0$,
\begin{equation}\label{Hp}
  [-\nu_Fk_x\tilde{\sigma}_y+(m_0-m_1k_x^2+M_z)\tilde{\sigma}_z]\zeta_+(x) = 0.
\end{equation}
It can be checked that $H_+(k_y=0)$ and $\tilde{\sigma}_x$ satisfy the anticommutation relation $\{H_+(k_y=0),\tilde{\sigma}_x\}=0$.
Thus, the zero-energy eigenstate is the simultaneous eigenstate of $H_+(k_y=0)$ and $\tilde{\sigma}_x$.
Consider the ansatz $\zeta_+ (x)=\eta_+^s(x)\chi_x^s$, where $\tilde{\sigma}_x\chi_x^s=s\chi_x^s$, $s=\pm1$, and $\chi_x^s=\frac{1}{\sqrt{2}} \binom{1}{s}$, we have
\begin{eqnarray}
  (s\nu_F\partial_x+m_0+m_1\partial_x^2+M_z)\eta_+^s(x) &=& 0 .
\end{eqnarray}
In order to solve the differential equation, we use instead of $x$ the variable $u=(1+e^{2x/\delta})^{-1}$.\cite{Flugges} With
\begin{eqnarray*}
  \frac{d}{dx}&=&-\frac{2}{\delta}u(1-u)\frac{d}{du},\\
  \frac{d^2}{dx^2}&=&\frac{d^2u}{dx^2}\frac{d}{du}+\left(\frac{du}{dx}\right)^2\frac{d^2}{du^2}\\
  &=& \left(\frac{2}{\delta}\right)^2u(1-u)(1-2u)\frac{d}{du}+\left(\frac{2}{\delta}\right)^2[u(1-u)]^2\frac{d^2}{du^2},\\
  \tanh\frac{x}{\delta}&=&\frac{e^{\frac{x}{\delta}}-e^{-\frac{x}{\delta}}}{e^{\frac{x}{\delta}}+e^{-\frac{x}{\delta}}}=1-\frac{2e^{-\frac{x}{\delta}}}{e^{\frac{x}{\delta}}+e^{-\frac{x}{\delta}}}=1-2u,
\end{eqnarray*}
we arrive at,
\begin{eqnarray}
  \left[\left(\frac{2}{\delta}\right)^2 u(1-u)(1-2u)\frac{d}{du}+\left(\frac{2}{\delta}\right)^2u^2(1-u)^2\frac{d^2}{du^2}
  -\frac{2s\nu_F}{m_1\delta}u(1-u)\frac{d}{du}+\frac{m_0}{m_1}-\frac{M}{m_1}(1-2u)\right]\eta_+^s(u) &=& 0,\\
  \left[u(1-u)\frac{d^2}{du^2}+(1-2u+\lambda_1)\frac{d}{du}+\frac{\lambda_2}{u(1-u)}+\frac{\lambda_3(1-2u)}{u(1-u)}\right]\eta_+^s(u) &=&0,\label{dif1}
\end{eqnarray}
with $\lambda_1=-\frac{\delta s\nu_F}{2m_1}$, $\lambda_2=\frac{\delta^2m_0}{4m_1}$, and $\lambda_3=-\frac{\delta^2M}{4m_1}$.
This equation has poles at $u=0,1,\infty$ and therefore leads to hypergeometric solutions.
Let's set
\begin{equation}
  \eta_+^s(u)=u^\alpha(1-u)^\beta f_+^s(u),
\end{equation}
with
\begin{eqnarray}
  \alpha &=& \frac{\delta}{2}\frac{\sqrt{\nu_F^2-4m_1(m_0-M)}-\nu_F}{2m_1},\label{alpha} \\
  \beta &=& \frac{\delta}{2}\frac{\nu_F-\sqrt{\nu_F^2-4m_1(m_0+M)}}{2m_1},\label{beta}
\end{eqnarray}
we get
\begin{eqnarray*}
 \frac{d\eta_+^s}{du} &=& \alpha u^{\alpha-1}(1-u)^\beta f(u)-\beta u^\alpha(1-u)^{\beta-1}f(u)+u^\alpha(1-u)^\beta f^\prime(u), \\
 \frac{d^2\eta_+^s}{du^2} &=& \alpha(\alpha-1)u^{\alpha-2}(1-u)^\beta f(u)-\alpha \beta u^{\alpha-1}(1-u)^{\beta-1}f(u)+\alpha u^{\alpha-1}(1-u)^\beta f^\prime(u)\\
 &&-\alpha \beta u^{\alpha-1}(1-u)^{\beta-1}f(u)+\beta(\beta-1)u^\alpha(1-u)^{\beta-2}f(u)-\beta u^\alpha(1-u)^{\beta-1}f^\prime(u)\\
 &&+\alpha u^{\alpha-1}(1-u)^\beta f^\prime(u)-\beta u^\alpha(1-u)^{\beta-1}f^\prime(u)+u^\alpha(1-u)^\beta f^{\prime\prime}(u),\\
 (1-2u)\frac{d\eta_+^s}{du} &=&[\alpha u^{\alpha-1}(1-u)^{\beta+1}-\beta u^\alpha(1-u)^\beta-\alpha u^\alpha(1-u)^\beta+\beta u^{\alpha+1}(1-u)^{\beta-1}]f(u)\\
 &&+[u^\alpha(1-u)^{\beta+1}-u^{\alpha+1}(1-u)^\beta]f^\prime(u),\\
 \lambda_2\frac{\eta_+^s}{u(1-u)}&=&\lambda_2u^{\alpha-1}(1-u)^{\beta-1}f(u),\\
 \lambda_3\frac{(1-2u)\eta_+^s}{u(1-u)}&=&\lambda_3(\frac{1}{u}-\frac{1}{1-u})\eta_+^s\\
 &=&\lambda_3u^{\alpha-1}(1-u)^\beta f(u)-\lambda_3u^\alpha(1-u)^{\beta-1}f(u).
 \end{eqnarray*}
Then substituting these expressions into Eq.(\ref{dif1}),
we arrive at the Gaussian equation
\begin{eqnarray}\label{gaussian}
  u(1-u)f^{\prime\prime}(u)+[(2\alpha+1+\lambda_1) -(2\alpha+2\beta+2)u]f^\prime(u)-(\alpha+\beta)(\alpha+\beta+1)f(u)=0.
\end{eqnarray}
In the derivation of Eq.(\ref{gaussian}), we have used identity $\alpha^2+\lambda_1 \alpha+\lambda_3+\lambda_2 =0$ and $\beta^2-\lambda_1\beta-\lambda_3+\lambda_2 =0$.
Then Eq.(\ref{gaussian}) has the special solution
\begin{equation}\label{hypo}
  f(u)=\sqrt{2} K_1\ _{2}F_1(\alpha+\beta,\alpha+\beta+1,2\alpha+1+\lambda_1;u),
\end{equation}
with the hypergeometric function $_{2}F_1$ and a normalized constant $K_1$.
To determine the value of $s$, we turn to the boundary conditions $\eta_+^s(-\infty)=0$ and $\eta_+^s(+\infty)=0$.

For the limit $x\rightarrow-\infty$ or $u\rightarrow 1$, $1-u = e^{2x/\delta}/(1+e^{2x/\delta})
 \simeq e^{2x/\delta}\rightarrow0$.
We apply the transformation rules for passing over from the argument $u$ to $1-u$ of the hypergeometric function,
\begin{eqnarray}
  &_{2}F_{1}&(\alpha+\beta,\alpha+\beta+1,2\alpha+1+\lambda_1;u)\nonumber\\
  &=&\frac{\Gamma(2\alpha+1+\lambda_1)\Gamma(\lambda_1-2\beta)}{\Gamma(\alpha-\beta+1+\lambda_1)\Gamma(\alpha-\beta+\lambda_1)}\times
  \ _{2}F_1(\alpha+\beta,\alpha+\beta+1,2\beta+1-\lambda_1;1-u)\nonumber\\
 &+&\frac{\Gamma(2\alpha+1+\lambda_1)\Gamma(2\beta-\lambda_1)}{\Gamma(\alpha+\beta)\Gamma(\alpha+\beta+1)}(1-u)^{\lambda_1-2\beta}\times
  \ _{2}F_1(\alpha-\beta+\lambda_1,\alpha-\beta+\lambda_1+1,-2\beta+1+\lambda_1;1-u)\nonumber.\\
\end{eqnarray}
With $1-u=e^{2x/\delta}$ and $_{2}F_1(0)=1$, this leads to
\begin{eqnarray}\label{minfty3}
  \eta_+^s(x)&=& \sqrt{2} K_1u^\alpha(1-u)^\beta \left\{\frac{\Gamma(2\alpha+1+\lambda_1)\Gamma(\lambda_1-2\beta)}{\Gamma(\alpha-\beta+1+\lambda_1)\Gamma(\alpha-\beta+\lambda_1)}
  +\frac{\Gamma(2\alpha+1+\lambda_1)\Gamma(2\beta-\lambda_1)}{\Gamma(\alpha+\beta)\Gamma(\alpha+\beta+1)}(1-u)^{\lambda_1-2\beta}\right\}\nonumber\\
 &\rightarrow&
 \sqrt{2} K_1\left\{\frac{\Gamma(2\alpha+1+\lambda_1)\Gamma(\lambda_1-2\beta)}{\Gamma(\alpha-\beta+1+\lambda_1)\Gamma(\alpha-\beta+\lambda_1)}e^{2\beta x/\delta}
 +\frac{\Gamma(2\alpha+1+\lambda_1)\Gamma(2\beta-\lambda_1)}{\Gamma(\alpha+\beta)\Gamma(\alpha+\beta+1)}e^{2(\lambda_1-\beta)x/\delta}\right\}.
\end{eqnarray}
The boundary condition $\eta_+^s(-\infty)=0$ implies that $\beta<\lambda_1$ and $\beta>0$.
From Eq.(\ref{beta}), one can see that the condition $\beta>0$ is satisfied always.
From the condition $\beta<\lambda_1$ and $\lambda_1=-\frac{\delta s\nu_F}{2m_1}$,
$s$ can only take $-1$.
On the other hand, for the limit $x\rightarrow+\infty$ or $u\simeq e^{-2x/\delta}\rightarrow0$, the solution (\ref{hypo}) becomes $f(0)=\sqrt{2} K_1$ or
\begin{equation*}\label{minfty}
  \eta_+^{-}(x)\rightarrow \sqrt{2} K_1u^\alpha \simeq \sqrt{2} K_1e^{-2\alpha x/\delta}.
\end{equation*}
The boundary condition $\eta_+^s(\infty)=0$ implies that $\alpha>0$.
From Eq.(\ref{alpha}), this condition is satisfied at $M>m_0$.
Finally, we obtain the wavefunction of zero energy for $H_+$,
\begin{equation}\label{solution1}
  \zeta_+(x)= \eta_+^-(x)\chi_x^- =
   K_1u^{\alpha}(1-u)^{\beta}\ _{2}F_1(\alpha+\beta,\alpha+\beta+1,2\alpha+1+\frac{\delta\nu_F}{2m_1};u)(1,-1)^T.
\end{equation}

Next, we solve the eigenequation $H_-\zeta_-(x)=E\zeta_-(x)$ for $k_y=0$ and $E=0$,
\begin{equation}\label{Hm}
  [\nu_Fk_x\tilde{\sigma}_y+(m_0-m_1k_x^2-M_z)\tilde{\sigma}_z]\zeta_-(x) = 0.
\end{equation}
In order to solve this differential equation, we use instead of $x$ the variable $g=(1+e^{-2x/\delta})^{-1}$ and considering the ansatz $\zeta_- (x)=\eta_-^s(x)\chi_x^s$, the Eq.(\ref{Hm}) becomes
\begin{eqnarray}
 \left[g(1-g)\frac{d^2}{dg^2}+(1-2g+\lambda_1)\frac{d}{dg}+\frac{\lambda_2}{g(1-g)}+\frac{\lambda_3(1-2g)}{g(1-g)}\right]\eta_-^s(g) &=&0,\label{dif2}
\end{eqnarray}
which has the same form as Eq.(\ref{dif1}). This allows us to reuse the previous results. To satisfy the boundary conditions $\eta_-^s(-\infty)=0$ and $\eta_-^s(+\infty)=0$,
we can obtain the wave function of zero energy for $H_-$,
\begin{equation}\label{solution2}
  \zeta_-(x)=\eta_-^-(x)\chi_x^- = K_2g^\alpha(1-g)^\beta \ _{2}F_1(\alpha+\beta,\alpha+\beta+1,2\beta+1+\frac{\delta\nu_F}{2m_1};g)(1,-1)^T
\end{equation}
with a normalized constant $K_2$.

Figure S1 displays the distributions of the probability density and the expectation value of $\tilde{\sigma}_x,\tilde{\sigma}_y,\tilde{\sigma}_z$ in the bound state $\zeta_\pm$ compared with numerical calculation. Both states are distributed around the DW, and only $\tilde{\sigma}_x$ is non-vanishing and negative which is consist with $s=-1$.
Moreover, the analytical results are well consistent with numerical results.

\begin{figure}
\includegraphics[scale=0.5]{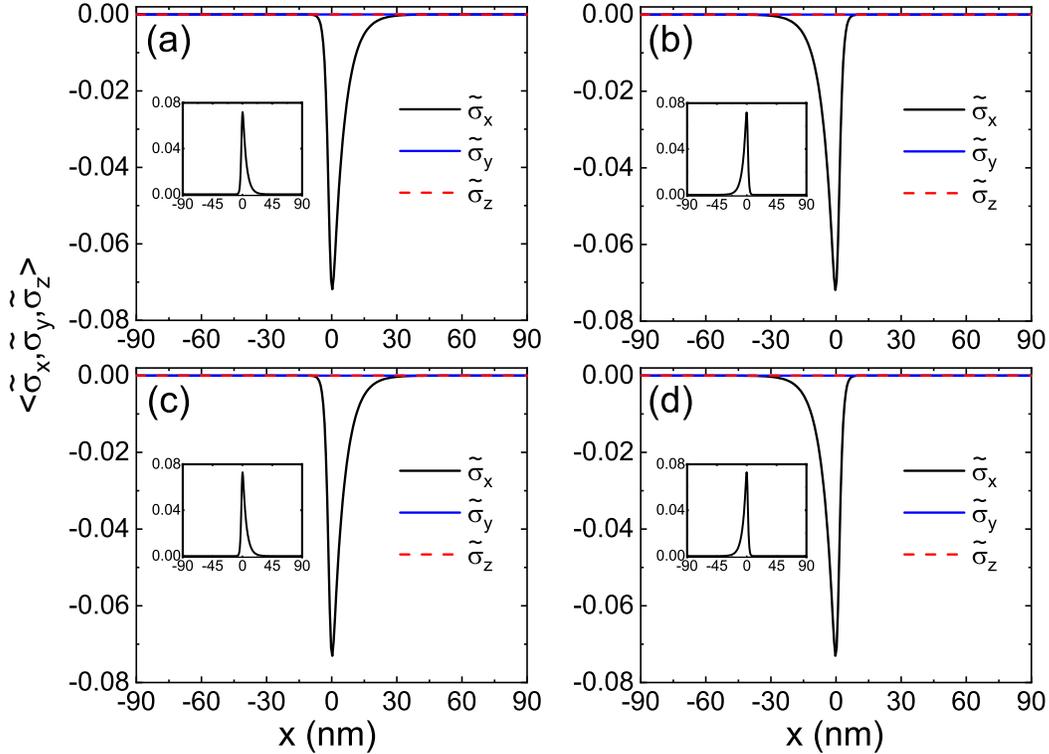}
\caption{Distribution of expectation value of $\tilde{\sigma}_x,\tilde{\sigma}_y,\tilde{\sigma}_z$ in the bound state at the DW solved from (a) $H_+$  and (b) $H_-$ with momentum $k_y=0$ by numerical calculation.
(c) and (d) are the analytic results from $\zeta_+(x)$ and $\zeta_-(x)$ in Eqs.(\ref{solution1}) and (\ref{solution2}).
The insets display probability density of the bound states.}
\end{figure}

\end{widetext}

\end{document}